\documentclass[sigconf]{acmart}

\AtBeginDocument{%
  }

\copyrightyear{2025}
\acmYear{2025}
\setcopyright{rightsretained}
\acmConference[CHI EA '25]{Extended Abstracts of the CHI Conference on Human Factors in Computing Systems}{April 26-May 1, 2025}{Yokohama, Japan}
\acmBooktitle{Extended Abstracts of the CHI Conference on Human Factors in Computing Systems (CHI EA '25), April 26-May 1, 2025, Yokohama, Japan}\acmDOI{10.1145/3706599.3719915}
\acmISBN{979-8-4007-1395-8/2025/04}

\usepackage{algorithmic}
\usepackage{algorithm}
\usepackage{graphicx}    
\usepackage{subcaption}   
\usepackage{booktabs}
\usepackage{enumitem}
\usepackage{stfloats}

\begin{document}

\title{iMedic: Towards Smartphone-based Self-Auscultation Tool for AI-Powered Pediatric Respiratory Assessment}

\author{Seung Gyu Jeong}
\email{wa3229433@gmail.com}
\affiliation{%
  \institution{Seoul National University of Science and Technology}
  \city{Seoul}
  \country{South Korea}}

\author{Sungwoo Nam}
\email{namsw@woorisoa.co.kr}
\affiliation{%
  \institution{Woorisoa Children's Hospital}
  \city{Seoul}
  \country{South Korea}
}

\author{Seongkwan Jung}
\email{unmouse@gmail.com}
\affiliation{%
 \institution{Woorisoa Children's Hospital}
 \city{Seoul}
 \country{South Korea}}

\author{Seong-Eun Kim}
\email{sekim@seoultech.ac.kr}
\affiliation{%
  \institution{Seoul National University of Science and Technology}
  \city{Seoul}
  \country{South Korea}}

\renewcommand{\shortauthors}{Jeong et al.}

\begin{abstract}
Respiratory auscultation is crucial for early detection of pediatric pneumonia, a condition that can quickly worsen without timely intervention. In areas with limited physician access, effective auscultation is challenging. We present a smartphone-based system that leverages built-in microphones and advanced deep learning algorithms to detect abnormal respiratory sounds indicative of pneumonia risk. Our end-to-end deep learning framework employs domain generalization to integrate a large electronic stethoscope dataset with a smaller smartphone-derived dataset, enabling robust feature learning for accurate respiratory assessments without expensive equipment. The accompanying mobile application guides caregivers in collecting high-quality lung sound samples and provides immediate feedback on potential pneumonia risks. User studies show strong classification performance and high acceptance, demonstrating the system’s ability to facilitate proactive interventions and reduce preventable childhood pneumonia deaths. By seamlessly integrating into ubiquitous smartphones, this approach offers a promising avenue for more equitable and comprehensive remote pediatric care.
\end{abstract}

\begin{teaserfigure}
  \includegraphics[width=\textwidth]{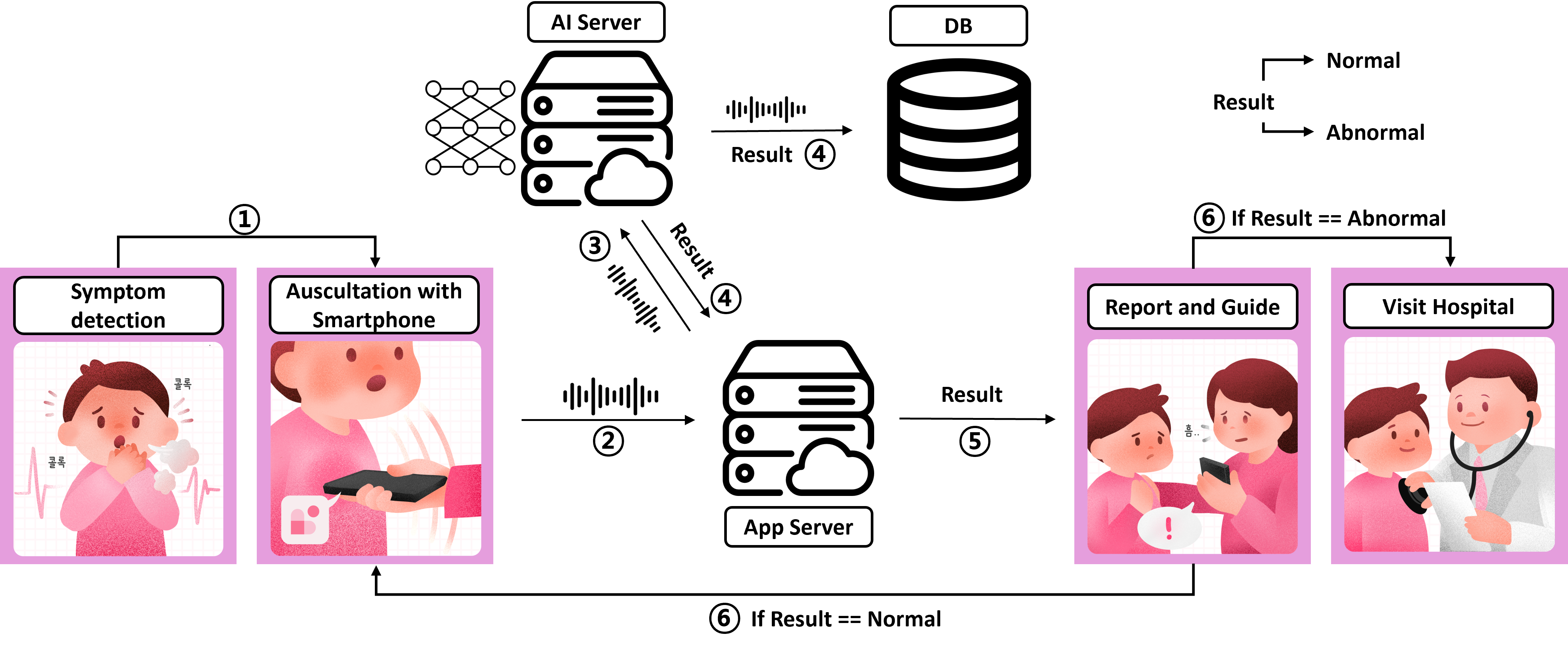}
  \caption{An overview of the proposed smartphone-based lung sound analysis pipeline. A user experiencing respiratory symptoms records lung sounds via the iMedic mobile application using the smartphone's built-in microphone. The recorded data is then transmitted to an AI server, where our deep learning model analyzes and classifies the lung sounds as normal or abnormal. The results are stored in a database to continuously improve the AI model, and the outcome is relayed back to the user through the app, guiding them to seek hospital care if necessary. }
  \Description{This figure shows a step-by-step schematic of the pipeline for analyzing lung sounds via a smartphone. On the far left, there is a simplified icon or illustration of a person (representing a pediatric patient or user) holding a smartphone against their chest. The smartphone’s built-in microphone is used to record lung sounds. Arrows lead to the central component labeled “AI Server,” which processes the recorded audio. Here, a deep learning model classifies the lung sounds as “normal” or “abnormal.”
  Another arrow leads from the AI Server to a “DB” (database) component, where results are stored and used for continuously improving the AI model. Finally, the outcome is sent back to the user’s smartphone (on the right side of the figure), prompting guidance such as whether to visit a hospital. Overall, the flow is user → AI server → database → back to user. The primary goal is to illustrate how smartphone-collected audio is processed by an AI model, with the result and recommendation returned to the user in real time.}
  \label{fig:teaser}
\end{teaserfigure}


\maketitle

\section{INTRODUCTION}

Respiratory diseases are among the leading causes of morbidity and mortality worldwide, affecting millions of people each year \cite{ref1,ref2,ref3,ref4}. Asthma, chronic obstructive pulmonary disease (COPD), and lower respiratory tract infections significantly contribute to medical costs and adversely affect the quality of life \cite{ref5}. Children and infants, in particular, are highly vulnerable due to their underdeveloped immune systems and narrower airways, facing long-term health risks and economic challenges \cite{ref6}. This underscores the critical need for effective and scalable diagnostic tools that can operate across diverse settings and resource-constrained environments.

Lung auscultation is widely recognized as a cornerstone of respiratory disease diagnosis and is a non-invasive and cost-effective method \cite{ref7}. However, traditional stethoscopes depend heavily on the auditory expertise of medical professionals, limiting their utility in remote or underserved areas. The COVID-19 pandemic further highlighted the importance of telemedicine, emphasizing the need for innovative diagnostic solutions that overcome geographical and resource barriers \cite{ref9, ref10, ref11, ref16}.

Recent advancements in electronic stethoscopes have enabled the collection of extensive datasets, leading to significant progress in AI-based lung sound classification \cite{nguyen2022lung,ma2019lungbrn,rocha2018alpha,bae2023patch}. However, these specialized devices are relatively expensive and have low adoption rates in resource-limited settings.

Smartphones, equipped with high-quality microphones, widespread connectivity, and extensive adoption, have emerged as an ideal platform for real-time health monitoring and diagnostics \cite{ref15, ref17}. Previous studies have demonstrated that smartphones can capture behavioral and physiological data for healthcare applications, particularly in resource-constrained settings \cite{ref13, ref18, ref20}. More recently, applications have been developed to measure lung sounds using smartphone microphones\cite{ref14}, enabling parents to record their children’s lung sounds at home and analyze the quality of the collected data\cite{ref21}. 

However, these studies have predominantly focused on demonstrating the feasibility of lung sound acquisition and analysis using smartphones. Healthcare professionals compared the lung sound data collected via smartphones with those acquired using an electronic stethoscope, confirming that the smartphone recordings were of sufficiently high quality to distinguish abnormal sounds. Despite promising findings, most prior work has not incorporated AI-driven methods for automatically classifying abnormal lung sounds. Although smartphone-based lung sound acquisition is feasible in non-clinical environments, robust AI solutions for smartphone-based abnormal sound detection remain largely unexplored.

In this study, we propose \textit{iMedic}, an end-to-end smartphone-based self-auscultation application for AI-powered pediatric respiratory assessment that relies solely on built-in smartphone microphones. The first “i” in \textit{iMedic} is deliberately chosen to evoke the Korean term for "child," underscoring our focus on pediatric care and embodying the role of a pediatrician. By leveraging the built-in microphones of smartphones, \textit{iMedic} guides caregivers through an intuitive process for correctly positioning the device and recording high-quality lung sounds. With clear visual instructions and real-time feedback, the app not only
facilitates precise data collection but also enhances user engagement and trust in the diagnostic process.

However, developing a high-performance AI system based on smartphone recordings for respiratory sound classification is a significant challenging. The high volume of annotated data from electronic stethoscopes has driven substantial progress in deep-learning models for abnormal lung sound classification \cite{ref23,ma2019lungbrn,bae2023patch}. Data acquired directly from smartphone microphones are still limited, hindering the achievement of reliable and accurate diagnostic performance. To overcome this shortage of labeled smartphone recordings, we propose a domain generalization framework that learns latent features by integrating a large-scale electronic stethoscope dataset with a smaller corpus of smartphone-collected lung sounds. This deep-learning framework enables the model to learn robust and discriminative feature representations across heterogeneous data sources, achieving accurate respiratory assessments without the need for additional costly equipment.

Together, these technological innovations establish a virtuous cycle: as more lung sound data are collected through our accessible platform, the deep learning model continuously improves its accuracy, which in turn boosts user confidence and engagement. By promoting user-friendly interfaces and practical applications in daily life, our approach strengthens human-computer interaction (HCI) and demonstrates tangible benefits in real-world scenarios. This study highlights the innovative application of smartphone-based lung sound classification and user-centric mobile application design, enabling early diagnosis and management of respiratory diseases in resource-limited contexts. Ultimately, our work aims to reduce respiratory disease-related morbidity and mortality, contributing to a decrease in global health disparities.

\begin{figure*}[t!]
  \centering
  \includegraphics[width=\textwidth]{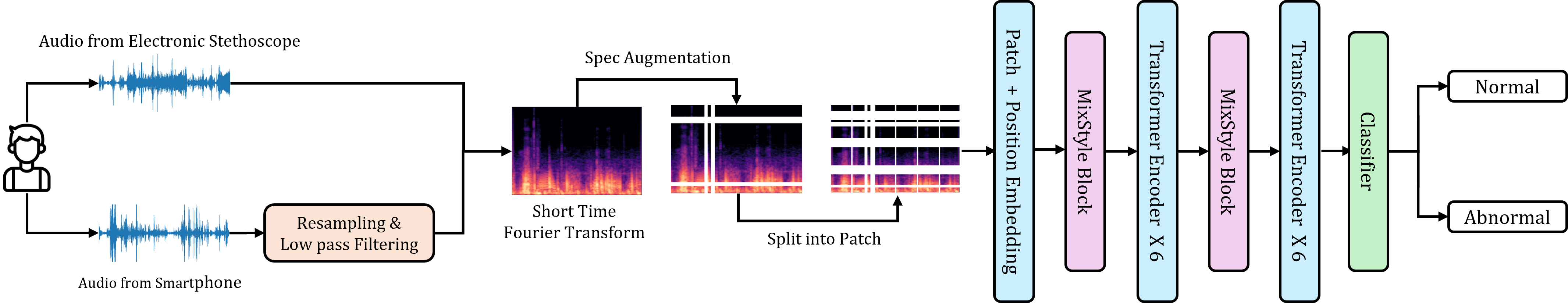}
  \caption{Overview of the deep learning model architecture, including the preprocessing, AST backbone, and MixStyle block.}
  \label{fig:deeplearning}
  \Description{This figure is a left-to-right flow diagram illustrating how recorded lung sounds pass through a deep learning pipeline:
  Input Audio (Left Side): Two separate audio waveforms (blue wave plots) represent lung sound recordings from either a smartphone or digital stethoscope. An icon of a person’s head and shoulders is also shown, symbolizing the user or patient.
“Resampling & Low pass Filtering” Block: The next box indicates that the audio is resampled to a uniform sample rate and passes through a low-pass filter, removing high-frequency noise.
Spectrogram Conversion: The filtered signals are transformed into spectrograms—colored heatmap-like images indicating frequency and intensity over time.
Patch + Position Embedding: The spectrograms are divided into small rectangular patches (shown by a grid overlay). Each patch is enriched with positional information before it flows through subsequent blocks.
MixStyle Block (Pink Box): A domain generalization method is applied here, mixing feature statistics (mean and standard deviation) across different recording devices but keeping the same class labels. This helps the model focus on clinically relevant sound features rather than device-specific noise.
Transformer Encoder x 6 (Light Blue Boxes): Multiple transformer encoder layers process these embedded patches to extract higher-level features crucial for classifying lung sounds.
Classifier (Green Box): Finally, a classification head uses the extracted features to label the lung sounds as normal, crackle, wheeze, or both (crackle + wheeze).
This pipeline helps mitigate device-specific discrepancies and achieves robust lung sound classification in real-world settings.}
\end{figure*}

\begin{table}[t!]
  \centering
  \caption{Data Distribution and Patient Demographics}
  \resizebox{0.45\textwidth}{!}{%
    \begin{tabular}{lcc}
      \toprule
      \textbf{Metric} & \textbf{Electronic Stethoscope} & \textbf{Smartphone} \\
      \midrule
      \textbf{Demographics} & & \\
      Total Patients & 707 & 195 \\
      Male/Female (\%) & 55.2\% / 44.8\% & 48.2\% / 51.8\% \\
      Age (months) & 31.7 ± 23.9 & 35.7 ± 20.9 \\
      \midrule
      \textbf{Data Distribution} & & \\
      Total Samples & 13,023 & 1,298 \\
      Normal & 6,955 & 888 \\
      Crackle & 2,296 & 182 \\
      Wheeze & 1,901 & 96 \\
      Both & 1,871 & 132 \\
      Recording Duration (seconds) & 3.06 ± 0.03 & 3.07 ± 0.03 \\
      Sampling Rate (Hz) & 4,000 & 48,000 \\
      \bottomrule
    \end{tabular}
  }
  \label{tab:data_and_demographics}
\end{table}

\section{METHOD}
\subsection{Data Collection}

This study was conducted in collaboration with Children's Hospital in Seoul, South Korea, and was approved by the Institutional Review Board of the university (IRB number: 2021-0017-02). Data were collected from pediatric patients aged 0–6 years, from October 27, 2021, to December 31, 2023. As part of standard clinical examinations, lung sounds were initially recorded using an electronic stethoscope. For patients (or guardians) who provided additional consent, subsequent recordings were made using a smartphone's built-in microphone immediately after the stethoscope recordings. The detailed patient demographics and data distribution are summarized in Table~\ref{tab:data_and_demographics}.  

\textbf{1) Electronic Stethoscope Data Collection---}  
We collected stethoscope recordings from 707 pediatric patients using a 3M™ Littmann® CORE Digital Stethoscope at four standard auscultation sites: right upper lobe (RUL), left upper lobe (LUL), left lower lobe (LLL), and right lower lobe (RLL)\cite{charbonneau2000basic,rossi2000environmental}. Each recording had an average duration of 3.06 seconds (SD = 0.03 seconds) and was sampled at 4,000 Hz. In total, 13,023 samples were collected and categorized into four classes: 6,955 normal lung sounds (53.40\%), 2,296 crackle sounds (17.63\%), 1,901 wheeze sounds (14.60\%), and 1,871 simultaneous crackle and wheeze (``both'') sounds (14.37\%).

\textbf{2) Smartphone Data Collection---}  
We collected smartphone recordings from 195 pediatric patients using an iPhone 13 immediately after the electronic stethoscope recordings. To ensure consistency, the smartphone’s microphone was placed at the same four auscultation sites (RUL, LUL, LLL, RLL). A total of 195 pediatric patients consented to this additional recording process. Each recording had an average duration of 3.07 seconds (SD = 0.03 seconds) and was sampled at 48,000 Hz. A total of 1,298 samples were collected and categorized into four classes: 888 normal lung sounds (68.41\%), 182 crackle sounds (14.02\%), 96 wheeze sounds (7.39\%), and 132 simultaneous crackle and wheeze (``both'') sounds (10.17\%).

\textbf{3) Data Quality Assurance and Labeling---}  
All recordings underwent in rigorous quality control process. Two independent clinicians reviewed the recordings to ensure data quality, excluding samples with excessive background noise, unclear lung sounds, or other artifacts. The remaining recordings were manually labeled into one of four categories: normal, crackle, wheeze, or both. This four-class scheme resulted in a significant class imbalance, which may hinder deep learning models by biasing training toward majority classes and increasing the risk of overfitting. To address these issues and to develop a smartphone-based screening tool for non-specialists, we simplified the labels into a binary classification of \textit{normal} and \textit{abnormal}. Specifically, we merged \textit{crackle}, \textit{wheeze}, and \textit{both} categories into a single \textit{abnormal} class to highlight any potential pathology in a user-friendly way. This meticulous review process ensured the reliability and clinical significance of the dataset for subsequent analysis.

\begin{figure*}[t!]
  \centering
  \includegraphics[width=\textwidth]{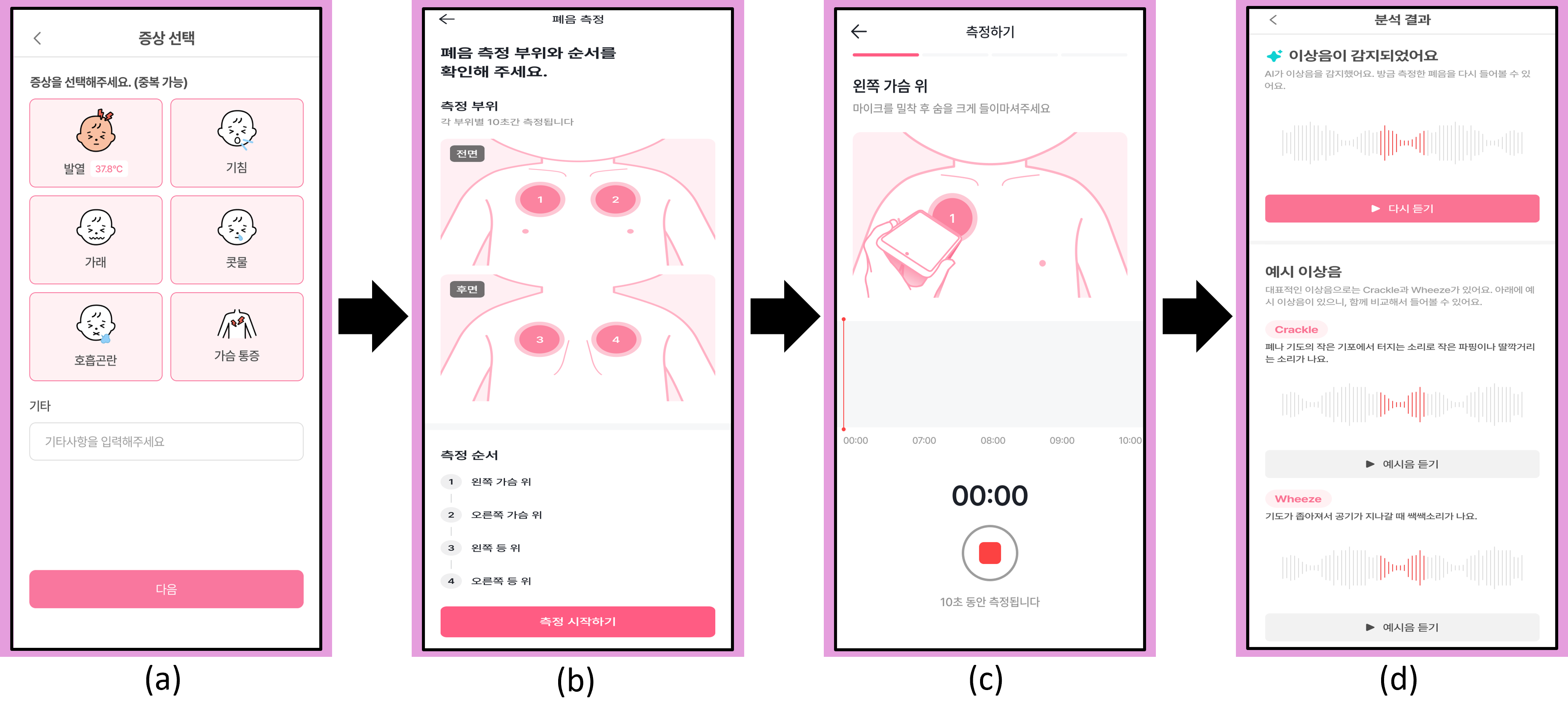}
  \caption{Screenshots of the application interface: (a) symptom selection screen, (b) recording site guidance screen, (c) real-time recording screen, (d) results display screen.}
  \label{fig:app_ui_horizontal}
  \Description{Four smartphone screen captures are displayed side by side, showing the main functionalities of the iMedic mobile app.
  
(a) Symptom Selection Screen:
The title at the top reads “Select Symptoms.”
Multiple pink buttons or cards labeled with common respiratory symptoms, such as “Fever (37.8°C),” “Cough,” “Sputum,” “Runny Nose,” “Breathing Difficulty,” and “Chest Pain.”
A text field labeled “Others” lets the user type additional notes.
A pink button labeled “Next” is located at the bottom.

(b) Recording Site Guidance Screen:
The title says, “Please confirm the auscultation sites and order.”
A simple torso illustration is shown, divided into four auscultation sites: two on the front (numbered 1 and 2) and two on the back (numbered 3 and 4).
Below the illustration, the text explains the order of measurement (e.g., “1) Left Chest Upper, 2) Right Chest Upper, 3) Left Back Upper, 4) Right Back Upper”).
A pink button labeled “Start Measurement” appears at the bottom.

(c) Real-Time Recording Screen:
The title at the top might say “Left Chest Upper,” indicating which site to record.
An illustration shows a person holding a smartphone against the left side of their chest, with instructions to take a deep breath.
A live waveform or audio graph is displayed at the bottom, updating in real time as the recording proceeds.
A timer (e.g., “00:00”) counts the seconds, and a note says, “Recording for 10 seconds.”
A circular red button indicates “Stop” or “Record,” allowing the user to pause or end the recording.

(d) Results Display Screen:
A heading might say “An Abnormal Sound was Detected” or “Your Results: Normal,” depending on the outcome.
A waveform diagram is shown, highlighting the captured audio segment.
Below this waveform, example abnormal sounds are labeled “Crackle” or “Wheeze,” with a button like “Play Example” to hear reference samples.
A pink button labeled “Listen Again” may allow playback of the user’s recording.
The screen also provides textual guidance on whether to visit a hospital or take further steps.}
\end{figure*}

\subsection{Deep Learning Model}
By integrating data from both electronic stethoscope and smartphone recordings, this study aimed to develop a robust and generalizable lung sound classification algorithm. The combination of these datasets facilitates the applications of deep learning techniques to advance emerging smartphone-based healthcare solutions.

In this study, we adopted the Audio Spectrogram Transformer (AST) as the backbone model due to its proven effectiveness in audio classification tasks \cite{gong2021ast,zaman2023survey}. To ensure compatibility between datasets, smartphone recordings were resampled to match the frequency range of stethoscope recordings, and low-pass filtering was applied to reduce high-frequency noise.

One of the primary challenges was addressing the domain differences between stethoscope and smartphone recordings. While stethoscope data is collected using specialized hardware optimized for capturing high-quality lung sound, smartphone data is more prone to environmental noise and variability in microphone positioning. These differences can degrade the quality of smartphone data, potentially reducing performance when training on combined datasets.

To address this challenge, we applied MixStyle, a domain generalization technique designed to minimize variability between domains while preserving essential class-specific features \cite{zhou2021domain,zhou2022domain,zhou2024mixstyle}. In our experiments, we modified MixStyle by mixing features based on the frequency axis and only combining sounds from different devices that share the same class label. Additionally, we performed mixing twice during the early and mid-training stages to achieve richer representations of diverse features.

MixStyle exchanges the mean and standard deviation between different domains under these conditions, guiding the model to focus on class-relevant features rather than device-specific noise. This process reduces device-specific inconsistencies, allowing the model to focus on clinically relevant features and accurately distinguish abnormal sounds regardless of the recording devices. The approach leverages high-quality stethoscope data to compensate for the limitations of smartphone recordings. Although smartphone data is smaller in size and more prone to noise, MixStyle facilitates the effective integration of both datasets, improving the model's generalization capabilities across devices. By addressing device-specific inconsistencies, this method helps build more robust and scalable solutions for smartphone-based healthcare applications.

\subsection{iMedic App for Pediatric Respiratory Assessment}
Building on the proposed technology, we developed iMedic, a mobile application designed to assist parents in easily recording their child’s lung sounds and receiving AI-based analyses for potential abnormalities. The iMedic app provides clear, step-by-step visual instructions for self-auscultation using a smartphone, guiding users to the correct sites for lung sound recording. With its intuitive and user-friendly interface, iMedic ensures effortless navigation even for users with no prior training. Additionally, the app securely stores recorded lung sound data, contributing to the continuous expansion of the lung sound database and enhancing the AI model’s classification accuracy over time.

\textbf{Step 1: Symptom Input---}iMedic begins by prompting users to select their child’s current symptoms, such as fever or cough, which are common indicators of respiratory illnesses like pneumonia \cite{cho2016respiratory}. This step provides valuable contextual information that can enhance the accuracy of AI-based classification by mapping symptoms to corresponding lung sounds. 

\textbf{Step 2: Guided Lung Sound Recording---}The app then guides users to record lung sounds from four specific sites on the chest and back: the upper right lobe, upper left lobe, lower right lobe, and lower left lobe. To minimize trial and error, a visual guide with clear illustrations shows the correct locations for recordings. Once the smartphone is correctly positioned at a designated site, the app records lung sounds for 10 seconds. On-screen visual cues ensure users follow the measurement steps properly, facilitating the collection of high-quality recordings essential for accurate AI-based assessment. 

\textbf{Step 3: AI-based Assessment---}The recorded lung sound data is transmitted to our AI server for assessment. The AI model evaluates lung sounds to detect potential abnormalities, such as wheezes, crackles, or both indicative of respiratory illnesses. Results are presented in an easy-to-understand format, categorizing the recordings as ``normal'' or recommending a medical consultation if abnormalities are detected.

By integrating an intuitive smartphone-based lung sound recording interface with AI-powered analysis, the iMedic app empowers parents to actively monitor their child's respiratory health and take timely, informed actions.

\begin{table*}
  \caption{Performance Comparison of Primary Experimental Setups. (Setup 1,2,3)}
  \label{tab:performance}
  \begin{tabular}{lcccc}
    \toprule
    Experiment Setup & SP (\%) & SE (\%) & Score (\%) & F1 Score \\
    \midrule
    AST (Smartphone Only) & 84.90 ± 13.74 & 75.90 ± 7.24 & 80.40 ± 4.25 & 0.74 ± 0.06 \\
    AST (Combined w/o MixStyle) & \textbf{91.60 ± 2.70} & 79.20 ± 6.23 & 85.40 ± 3.62 & 0.80 ± 0.05 \\
    AST (Combined w MixStyle) & 90.30 ± 4.69 & \textbf{83.30 ± 3.13} & \textbf{86.90 ± 3.18} & \textbf{0.82 ± 0.04} \\
    \bottomrule
  \end{tabular}
\end{table*}

\begin{table*}
  \caption{Performance Comparison of Supplementary Experimental Setups. (Setup 4,5)}
  \label{tab:stethoscope_vs_smartphone}
  \begin{tabular}{lcccc}
    \toprule
    Experiment Setup & SP (\%) & SE (\%) & Score (\%) & F1 Score \\
    \midrule
    AST (Stethoscope Only) & \textbf{86.83 ± 3.98} & \textbf{87.83 ± 2.81} & \textbf{87.33 ± 1.57} & \textbf{0.87 ± 0.02} \\
    AST (Tested on Smartphone) & 69.29 ± 8.74 & 86.30 ± 10.06 & 77.79 ± 4.54 & 0.68 ± 0.05 \\
    \bottomrule
  \end{tabular}
\end{table*}

\section{RESULTS}

\subsection{Experimental Setup}
We designed five experimental setups to validate the effectiveness of integrating a large-scale electronic stethoscope dataset with a smaller smartphone-collected dataset through our domain generalization technique to enhance smartphone-based lung sound classification performance. We conducted a five-fold cross-validation to provide a robust and unbiased evaluation of our model. In each fold, we randomly partitioned the data while maintaining a consistent ratio of normal to abnormal samples (i.e., stratification). This approach ensures that each fold reflects the overall class distribution, minimizing the risk of bias or overfitting that can arise when certain folds contain disproportionately high or low numbers of abnormal cases. This stratified strategy ultimately contributes to a more stable and representative assessment of the model's generalization capabilities. Appendix A (Table~\ref{tab:core_distribution}) provides a detailed summary of the data distribution across each fold and the total number of normal vs. abnormal samples in the training and test sets. The evaluation metrics included Sensitivity (Se), Specificity (Sp), the average of Se and Sp (Score), and the F1 score to account for class imbalance. \cite{rocha2018alpha}.

\textbf{Setup 1: Only Smartphone Dataset} In this setup, the model was trained and tested solely on smartphone data. The purpose was to assess the feasibility of using consumer-grade devices for lung sound classification without relying on specialized stethoscope data.

\textbf{Setup 2: Combined Dataset without Domain Generalization}
The model was trained on a combined dataset consisting of a large stethoscope dataset and a smaller smartphone dataset without applying any domain generalization techniques. The smartphone dataset was tested to evaluate the model's performance.

\textbf{Setup 3: Combined Dataset with Domain Generalization}
This setup incorporated our proposed MixStyle, a domain generalization technique, into the combined dataset of smartphone and stethoscope recordings. By leveraging the larger stethoscope dataset and reducing domain discrepancies, MixStyle enables the model to more efficiently extract representative features from the smaller smartphone dataset.

\textbf{Setup 4: Only Stethoscope Dataset}
This setup involved training and testing the model exclusively on high-quality stethoscope recordings. The objective was to benchmark the model's performance using a large, high-fidelity stethoscope dataset, thereby providing the baseline performance for comparison against the classification performance achieved with the considerably smaller smartphone dataset. 

\textbf{Setup 5: Stethoscope Trained, Smartphone Tested}
In this setup, the model was trained on stethoscope data and then tested on lung sounds recorded by smartphones. This scenario illustrates the challenges of domain discrepancies: a model trained solely on stethoscope data often struggles to interpret smartphone recordings accurately. The experiment emphasizes the AI algorithm's sensitivity to variations in recording devices and underscores the need for effective domain generalization techniques.

\subsection{Lung Sound Classification}
As shown in Table~\ref{tab:performance}, training the model solely with smartphone data yielded a Score of 80.4\%, while training and testing the model with stethoscope data achieved a higher Score of 87.33\%, highlighting the benefits of higher-quality and larger stethoscope datasets. Incorporating both stethoscope and smartphone data increased the Score to 85.4\%, along with relatively higher specificity and sensitivity. Moreover, applying our domain generalization technique (MixStyle) further improved performance on smartphone recordings by effectively mitigating device-specific discrepancies, achieving a Score of 86.9\% and an F1 score of 0.82. This result validates the effectiveness of MixStyle compared to a simple data combination approach. Conversely, when a model trained exclusively on stethoscope data was directly applied to smartphone data without any domain generalization, the score significantly dropped to 77.79\% compared to 87.33\% (Table~\ref{tab:stethoscope_vs_smartphone}). This clear performance decline underscores the adverse effects of domain mismatches between devices. Consequently, these findings underscore employing domain generalization methods to reduce device-specific inconsistencies and enhance the model’s applicability in real-world clinical environments.

\begin{table*}[t!]
\caption{Usability Testing Results (Likert Scale: 1 = Strongly Disagree, 5 = Strongly Agree)}
\label{tab:usability_results}
\centering
\begin{tabular}{p{0.75\linewidth}cc}
\toprule
\textbf{Survey Question} & \textbf{Mean Score} & \textbf{SD} \\
\midrule
1. I think that I would like to use this system frequently. & 4.3 & 0.7 \\
2. I found the system unnecessarily complex. & 1.7 & 0.8 \\
3. I thought the system was easy to use. & 4.3 & 0.6 \\
4. I think that I would need the support of a technical person to be able to use this system. & 1.9 & 0.8 \\
5. I found the various functions in this system were well integrated. & 4.1 & 0.7 \\
6. I thought there was too much inconsistency in this system. & 2.0 & 0.7 \\
7. I would imagine that most people would learn to use this system very quickly. & 4.0 & 0.7 \\
8. I found the system very cumbersome to use. & 1.8 & 0.8 \\
9. I felt very confident using the system. & 4.1 & 0.7 \\
10. I needed to learn a lot of things before I could get going with this system. & 1.9 & 0.7 \\
\bottomrule
\end{tabular}
\end{table*}

\subsection{Usability Testing of the Application}
To assess the usability of the iMedic app during lung sound collection, we conducted a structured usability survey. Participants were instructed to use the app to complete the full measurement process, which included symptom selection, guided microphone placement, and lung sound recording, and then respond to the System Usability Scale (SUS) questionnaire, consisting of 10 items evaluating various aspects of usability\cite{lewis2018system,bangor2008empirical,kaya2019usability,kortum2013usability}. In addition, we collected open-ended qualitative feedback to incorporate diverse perspectives. For this study, we recruited 12 volunteer parents, all in their mid-30s and experienced with smartphone usage, but without prior experience collecting lung sounds. These participants tested the iMedic app alongside their children, who were aged from 0 to 6 years.

The average SUS score was 78, which is considered high given that scores above 68 are generally regarded as above average usability \cite{lewis2018item}. Notably, questions such as "I think that I would like to use this system frequently." (Mean = 4.3, SD = 0.7) and "I thought the system was easy to use." (Mean = 4.3, SD = 0.6) received high scores, demonstrating that participants found the application both frequent-use friendly and easy to operate.

Open-ended feedback highlighted several potential improvements, such as incorporating real-time prompts or audio cues to confirm correct microphone placement and indicate the start or end of the recording process. One participant suggested, "It would be better if there was a simple vibration feedback when the recording starts." Another participant remarked, "Having a sound notification when the recording ends would be more convenient." These suggestions will be considered in future updates to further enhance the user experience.

\section{DISCUSSION}

\subsection{Challenges and Limitations}

While domain generalization techniques significantly enhanced the performance of smartphone data, the overall accuracy remains slightly lower compared to models exclusively trained on large-scale stethoscope datasets. This discrepancy reflects the inherent quality differences between smartphone and stethoscope recordings, suggesting that further refinement of domain generalization methods is necessary.

Additionally, the dataset used in this study was collected from a single hospital, limiting the diversity of environments and devices represented. The lack of data from various real-world settings, including different smartphone models and environmental conditions, may affect the model's generalizability. Expanding the dataset to include diverse scenarios and validating the system in broader contexts are critical next steps.

\subsection{Implications and Future Directions}
This study underscores the potential of smartphones as accessible diagnostic tools for respiratory health management, particularly in resource-limited settings. The integration of smartphone-based auscultation with advanced AI techniques provides a scalable solution to address global health disparities in respiratory diagnostics.

Future research will focus on expanding data collection to include diverse smartphone models, environments, and populations to enhance the model's robustness and generalizability. Efforts will also be directed toward refining the user experience by incorporating real-time feedback mechanisms and better mapping symptom data with lung sound recordings to improve classification accuracy. Furthermore, developing specialized models for age-specific or disease-specific predictions will be explored to better cater to individual patient needs.

\section{CONCLUSION}
This study presents a novel approach to lung sound classification using smartphone microphones and advanced domain generalization techniques. By integrating stethoscope and smartphone data, our system effectively bridges the gap between high-quality medical devices and widely accessible consumer technology. Our iMedic app enables non-expert users to easily collect lung sound data, showcasing the practicality and scalability of our approach. The findings highlight the potential of smartphone-based healthcare solutions to support the early diagnosis and management of respiratory diseases in resource-limited settings. By addressing global health disparities, this research paves the way for more equitable access to respiratory health monitoring and demonstrates the transformative potential of smartphones in digital healthcare.

\begin{acks}
This work was supported by the National Research Foundation of Korea (NRF) funded by the Ministry of Science and ICT (MSIT) of the Korean government under Grants RS-2023-00208492 and RS-2024-00422599.
\end{acks}

\bibliographystyle{ACM-Reference-Format}
\bibliography{ref}

\newpage
\appendix
\section*{Appendix}
\setcounter{table}{0}
\renewcommand{\thetable}{A\arabic{table}}
\section{Data Distribution}

\begin{table*}[b]
\centering
\caption{Data Distribution Across All Experiments (Exp 1--5) and Folds (1--5).
For each experiment, we list the number of normal vs.\ abnormal samples 
for stethoscope vs.\ smartphone in both the training and test sets, per fold.
``St.'' = Stethoscope, ``Ph.'' = Smartphone, ``Norm.'' = Normal, ``Abn.'' = Abnormal.}
\label{tab:core_distribution}
\begin{tabular}{c cccc cccc}
\toprule
\multicolumn{1}{c}{} & \multicolumn{4}{c}{\textbf{Train Set}} & \multicolumn{4}{c}{\textbf{Test Set}} \\
\cmidrule(lr){2-5}\cmidrule(lr){6-9}
\textbf{Fold} 
& \textbf{St.\ Norm.} & \textbf{St.\ Abn.} & \textbf{Ph.\ Norm.} & \textbf{Ph.\ Abn.} 
& \textbf{St.\ Norm.} & \textbf{St.\ Abn.} & \textbf{Ph.\ Norm.} & \textbf{Ph.\ Abn.}\\
\midrule

\multicolumn{9}{c}{\textbf{Exp 1: Smartphone Only}} \\
\midrule
1 & -  & -  & 703 & 336 & -  & -  & 185 & 74 \\
2 & -  & -  & 718 & 321 & -  & -  & 170 & 89 \\
3 & -  & -  & 718 & 319 & -  & -  & 170 & 91 \\
4 & -  & -  & 701 & 334 & -  & -  & 187 & 76 \\
5 & -  & -  & 712 & 330 & -  & -  & 176 & 80 \\
\midrule

\multicolumn{9}{c}{\textbf{Exp 2: Combined w/o MixStyle}} \\
\midrule
1 & 5641 & 4839 & 703 & 336 & - & - & 185 & 74 \\
2 & 5546 & 4937 & 718 & 321 & - & - & 170 & 89 \\
3 & 5514 & 4966 & 718 & 319 & - & - & 170 & 91 \\
4 & 5582 & 4908 & 701 & 334 & - & - & 187 & 76 \\
5 & 5513 & 4818 & 712 & 330 & - & - & 176 & 80 \\
\midrule

\multicolumn{9}{c}{\textbf{Exp 3: Combined w/ MixStyle}} \\
\midrule
1 & 5641 & 4839 & 703 & 336 & - & - & 185 & 74 \\
2 & 5546 & 4937 & 718 & 321 & - & - & 170 & 89 \\
3 & 5514 & 4966 & 718 & 319 & - & - & 170 & 91 \\
4 & 5582 & 4908 & 701 & 334 & - & - & 187 & 76 \\
5 & 5513 & 4818 & 712 & 330 & - & - & 176 & 80 \\
\midrule

\multicolumn{9}{c}{\textbf{Exp 4: Stethoscope Only}} \\
\midrule
1 & 5647 & 4840 & -  & -  & 1321 & 1298 & -  & -  \\
2 & 5558 & 4948 & -  & -  & 1410 & 1190 & -  & -  \\
3 & 5526 & 4967 & -  & -  & 1442 & 1171 & -  & -  \\
4 & 5594 & 4909 & -  & -  & 1374 & 1230 & -  & -  \\
5 & 5534 & 4818 & -  & -  & 1421 & 1250 & -  & -  \\
\midrule

\multicolumn{9}{c}{\textbf{Exp 5: Train on Stethoscope, Test on Smartphone}} \\
\midrule
1 & 5647 & 4840 & -  & -  & -  & -  & 185 & 74 \\
2 & 5558 & 4948 & -  & -  & -  & -  & 170 & 89  \\
3 & 5526 & 4967 & -  & -  & -  & -  & 170 & 91 \\
4 & 5594 & 4909 & -  & -  & -  & -  & 187 & 76 \\
5 & 5534 & 4818 & -  & -  & -  & -  & 176 & 80 \\
\bottomrule
\end{tabular}
\end{table*}

\end{document}